# Temporal Dynamic Synchronous Functional Brain Network for Schizophrenia Diagnosis and Lateralization Analysis

Cheng Zhu*, Ying Tan, Shuqi Yang*, Jiaqing Miao, Jiayi Zhu, Huan Huang, Dezhong Yao, and Cheng Luo

***Abstract*—The available evidence suggests that dynamic functional connectivity (dFC) can capture time-varying abnormalities in brain activity in resting-state cerebral functional magnetic resonance imaging (rs-fMRI) data and has a natural advantage in uncovering mechanisms of abnormal brain activity in schizophrenia(SZ) patients. Hence, an advanced dynamic brain network analysis model called the temporal brain category graph convolutional network (Temporal-BCGCN) was employed. Firstly, a unique dynamic brain network analysis module, DSF-BrainNet, was designed to construct dynamic synchronization features. Subsequently, a revolutionary graph convolution method, TemporalConv, was proposed, based on the synchronous temporal properties of feature. Finally, the first modular abnormal hemispherical lateralization test tool in deep learning based on rs-fMRI data, named CategoryPool, was proposed. This study was validated on COBRE and UCLA datasets and achieved 83.62% and 89.71% average accuracies, respectively, outperforming the baseline model and other state-of-the-art methods. The ablation results also demonstrate the advantages of TemporalConv over the traditional edge feature graph convolution approach and the improvement of CategoryPool over the classical graph pooling approach. Interestingly, this study showed that the lower order perceptual system and higher order network regions in the left hemisphere are more severely dysfunctional than in the right hemisphere in SZ and reaffirms the importance of the left medial superior frontal gyrus in SZ. Our core code is available at: https://github.com/swfen/Temporal-BCGCN.**

***Index Terms*—Dynamic functional connectivity, graph pooling, hemispherical lateralization, rs-fMRI, schizophrenia, temporal graph convolution.**

## I. INTRODUCTION

Currently, artificial intelligence methods that assist in diagnosing schizophrenia (SZ) and investigating hemispherical lateralization abnormalities are critical research topics in the field of clinical medicine for psychiatric disorders. Severe behavioral dysfunction is common among SZ patients, and the lifetime prevalence of approximately 1% [1] and a shorter life expectancy of approximately 15 years compared with the general population [2]. Concurrently, the diagnosis and treatment of SZ are arduous because a wide range of brain dysfunctions, in perception, thinking, emotion, and behavior, is involved. Resting-state functional magnetic resonance imaging (resting-state cerebral functional magnetic resonance imaging, rs-fMRI) [3], is a non-invasive medical imaging technique that measures fluctuations of blood oxygen level (BOLD) signals in various regions of interest (ROI) of the brain. rs-fMRI has been used to study abnormal functional connectivity patterns in SZ patients [4,5,6]. Although these studies have used computerized methods to analyze rs-fMRI data, there remains room to develop new computer-assisted diagnostic techniques for mental disorders and improve the usefulness of old methods. For example, making full use of rs-fMRI data to construct complex brain networks, improving feature extraction from the networks, and further investigating potential lateralization abnormalities in the patient's brain.

The size and complexity of brain network topologies pose challenges to understanding higher cognitive abnormalities in patients with psychiatric illnesses. Existing studies have shown that brain activity is always in a dynamic pattern of neural activity, and it is believed that dynamic functional connectivity (dFC) analysis can reveal temporal dynamics changes masked by static functional connectivity (sFC) [7, 8]. Therefore, the use of dFC to describe the state of brain network activity has become a research hotspot in complementary diagnosis and analysis of psychiatric disorders [9, 10, 11]. The dFC analysis uses a sliding window to slice the rs-fMRI data into multiple time segments, providing a new way to explore the dynamic mechanisms of brain activity further. Lin *et al.* [12] proposed a convolutional recurrent neural network (CRNN) for extracting high-level features of the dFC network and

*: Cheng Zhu and Shuqi Yang contributed equally to this work.. Correspondence to: Ying Tan, Jiaqing Miao, Cheng Luo.

This work was supported in part by the grant from National Key R&D Program of China 2018YFA0701400; in part by the National Natural Science Foundation of China 61933003, U2033271; in part by the Science and Technology Project in Sichuan of China under grant 2021ZYD0021, 2022NSFSC0530, 2022NSFSC0507; in part by the Sichuan Provincial Program of Traditional Chinese Medicine of China under grant 2021ZD017; and in part by the Fundamental Research Funds for the Central Universities of China, Southwest Minzu University, under grant 2021PTJS23. (Cheng Zhu and Shuqi Yang contributed equally to this work.) (Corresponding author: Ying Tan, Jiaqing Miao, Cheng Luo).

Cheng Zhu, Ying Tan, Shuqi Yang, Jiaqing Miao, Jiayi Zhu are with The Key Laboratory for Computer Systems of State Ethnic Affairs Commission, Southwest Minzu University, Chengdu, China. (e-mail: 190852112021@stu.swun.edu.cn; ty7499@swun.edu.cn; yangshuqi@stu.swun.edu.cn; jiaqing_miao@swun.edu.cn; zhujiayii0102@163.com).

Huan Huang, Dezhong Yao, Cheng Luo are with MOE Key Lab for Neuroinformation, High-Field Magnetic Resonance Brain Imaging Key Laboratory of Sichuan Province, The Clinical Hospital of Chengdu Brain Science Institute, University of Electronic Science and Technology of China, Chengdu 611731, China. (e-mail: huan_huang17@163.com; chengluo@uestc.edu.cn; dyao@uestc.edu.cn).

implement classification. Ramirez-Mahaluf *et al.* [13] found a temporal disorganization of dFC in the brains of patients with first-episode psychosis. Most studies have explored the dynamic patterns of change in brain networks using the dFC mechanisms but have not fully mined the implied synchronization trends between brain regions.

Graph neural networks (GNNs) extend deep learning (DL) methods to non-Euclidean domains to extract and aggregate feature information in graph-structured data more effectively, showing superior performance in the field of SZ-assisted diagnosis. Lei *et al.* [14] used a graph convolutional network (GCN) to extract features from rs-fMRI data to reveal abstract and complex relationships within brain networks and identify topological abnormalities in the functional networks of SZ patients. Chang *et al.*[15] applied a GNN to learn the topology of brain networks in the source space to better capture the complex relationships involved in the mismatch negative disorder mechanism in the electroencephalography of SZ patients. In this study, we found that most of the edge vector graph convolution methods, represented by Edge-Conditioned Convolution (ECConv) [16], disrupt the temporal construction of node and edge features during the convolution of traditional GNNs when applied to brain time series data. This disruption may degrade the classification performance of the model.

Some studies have found that SZ patients may have hemispheric lateralization abnormalities [4, 17] and disruptions to the mutual coordination of left and right hemisphere functions of the brain [18, 19]. Due to the essential role of pooling operations in the network structure, the use of pooling to verify brain lateralization abnormalities in SZ patients has become a topic of research interest. To exploit further the latent relationships among nodes, some studies clustered the nodes during pooling and then selected the nodes based on class clusters, where the selection was often data-driven. Yuan *et al.* [20] proposed a graph pooling technique that assigns nodes with similar features to the same class clusters and uses them for the prediction task of the graph structure. Gopinath *et al.* [21] introduced a learnable pooling strategy with a difference pooling technique to split the network into two separate paths, one for computing the latent features of nodes and the other for clustering the features into node class clusters. However, to verify the presence of abnormal lateralization in SZ patients, it is often necessary to divide the brain regions in the left and right hemispheres into two clusters according to prior knowledge. Existing graph clustering pooling strategies are limited in this case as they cannot target potential functional abnormalities in the left and right hemispheres of SZ patients.

To address these limitations, an advanced GCN improvement model based on dynamic brain network analysis, the temporal brain category graph convolutional network (Temporal-BCGCN), was proposed. Firstly, the dynamic synchronous functional brain network (DSF-BrainNet) with temporal dynamic synchronous properties was built using a sliding window technique to obtain information about nodes in brain regions and the interrelationships between nodes. Secondly, the proposed TemporalConv enables the convolution process to adapt to the dynamic synchronization rule in DSF-BrainNet. Finally, the pooling strategy CategoryPool was employed in the pooling layer of GCN to improve the accuracy of SZ diagnosis and investigate further hemispheric lateralization abnormalities of SZ patients.

The main contributions of this paper can be summarized as follows:

1) A unique dynamic brain network analysis module, DSF-BrainNet, is proposed. DSF-BrainNet brings new perspectives for understanding further the dynamic fluctuation patterns of various brain regions over time. This module was used as the input of the GCN.

2) A revolutionary convolutional method called TemporalConv is presented. TemporalConv convolves the features of the rs-fMRI data obtained from each time slice independently. The original concept of temporal convolution provides a new reference for subsequent researchers performing graph convolution on rs-fMRI and other temporal data.

3) A novel pooling strategy called CategoryPool is proposed. In rs-fMRI research, CategoryPool is the first modular hemispheric lateralization abnormalities test tool used for deep learning. This method has strong generalization and can be transferred to other medical research fields.

## II. DATASETS AND PREPROCESSING

Two datasets were selected for model evaluation in this study, the Center for Bio-medical Research Excellence (COBRE) Dataset [22] and University of California Los Angeles (UCLA) Dataset [23], with the following sources and corresponding preprocessing.

The COBRE Dataset was obtained from the COBRE, a free and publicly available database at the University of New Mexico Biomedical Research Center. This study selected rs-fMRI images of 112 subjects as simulation data, including 48 SZ patients and 64 healthy control (HC). Informed consent was obtained from each subject in accordance with the University of New Mexico Office of Human Research Protections. The UCLA Dataset was obtained from the UCLA Consortium for Neuropsychiatric Phenomics LA5c Study, which selected rs-fMRI images of 80 subjects as simulation data, including 41 SZ patients and 39 HC. All subjects gave written informed consent in accordance with procedures approved by the UCLA Institutional Review Board. In addition, subjects with a diagnosis of a neurological disorder, mental retardation, severe head trauma, substance abuse, or dependence within the previous 12 and six months were excluded from the aforementioned two datasets respectively.

The data were preprocessed using the DPABI toolbox [24] with the following main processes: 1) slice timing, 2) realignment, 3) normalization, 4) smoothing, and 5) detrending. In this study, the brain was divided into 116 ROIs using anatomical automatic labeling (AAL), and 26 ROIs in the cerebellar region were removed, resulting in 90 brain ROIs for the experiment.

## III. METHODS

### A. Overview of Temporal-BCGCN Architecture

In this study, an advanced Temporal-BCGCN model was proposed for computer-aided diagnosis and brain function lateralization analysis of SZ patients. The structural framework is illustrated in Fig. 1.

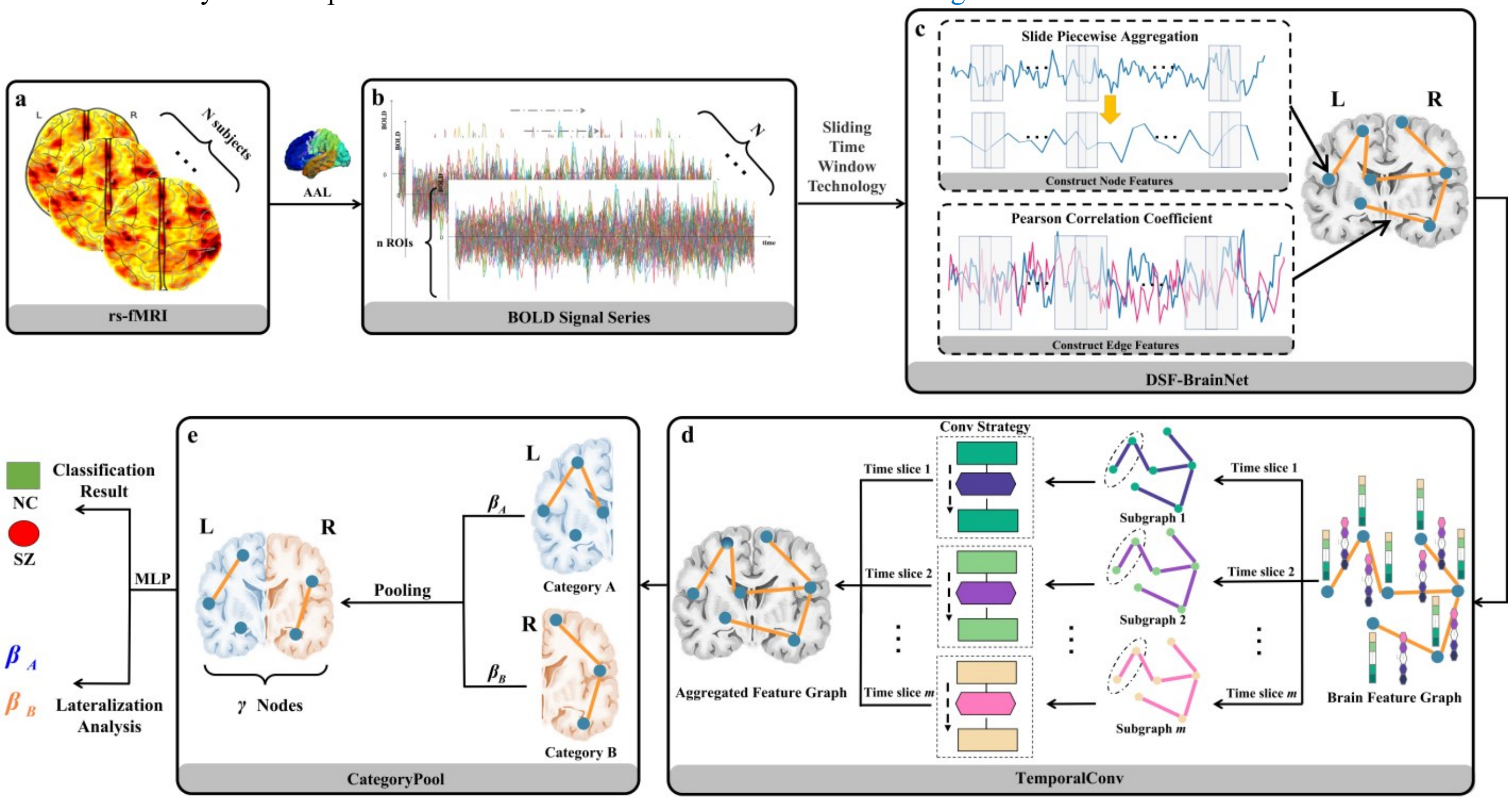

Fig. 1. Overall flowchart of the Temporal-BCGCN Architecture. (a) rs-fMRI. (b) BOLD Signal Series. (c) DSF-BrainNet. (d) TemporalConv. (e) CategoryPool.

Fig. 1(a) shows the brain's rs-fMRI data of $N$ subjects. The brain data of $N$ subjects are entered into the AAL template, and each subject is given the corresponding BOLD signal on each of the $n$ ROIs to obtain the $N$ sets of dataset in (b), where each set contains the BOLD signal sequence of the $n$ ROIs. The horizontal coordinates of the BOLD signal in Fig. 1(b) represent the time variation, the vertical coordinates represent the BOLD signal values, and the oscillating dashes in each color represent the specific BOLD signal values for each ROI at the corresponding time points. The BOLD signal in Fig. 1(b) is processed using a sliding window technique to obtain the time slices in Fig. 1 (c), and the node features and edge features are calculated for each time slice. Particularly, in Fig. 1(c), we propose the DSF-BrainNet module to construct a temporally synchronized feature brain network. First, we used the sliding window technique to divide the temporal signals of each subject into partially overlapping time slices. Then the signals of ROIs within each time slice were analyzed using the slide piecewise aggregation (SPA) technique to construct the graph node features. The Pearson correlation coefficient (PCC) between two ROIs within each time slice were calculated to construct the edge features of the graph network. Subsequently, the brain feature graphs constructed in Fig. 1(c) were fed into Fig. 1(d), the TemporalConv operation. We set up $m$ independent convolution paths according to the number of time slices, so that nodes and edge features under different time slices were signaled in independent paths, ensuring that the synchronous timing properties in DSF-BrainNet were not destroyed during the convolution process. In addition, the convolution module was run twice in succession. The aggregated feature graphs obtained in Fig. 1(d) were then fed into Fig. 1(e), where the left and right hemisphere nodes were set to Categories A and B, respectively. After determining the total number of nodes $\gamma$ to be retained in the pooling, the corresponding abnormal propensity parameters $\beta_A$ and $\beta_B$ were set to control the proportion of nodes retained in the left and right hemispheres, respectively. Finally, the pooled brain maps were fed into the multi-layer perceptron (MLP) to obtain the classification results. The optimal model values of $\beta_A$ and $\beta_B$ were determined through multiple experiments for further lateralization analysis.

### B. DSF-BrainNet

DSF-BrainNet is topologically composed of graph nodes and edge features. To reflect the fluctuation of brain activity over time, the graph nodes and edge features are organized as time series data, and they maintain a time-based synchronization property. They are defined as follows: $G=(V,E)$ denotes the undirected graph, where $V=\{v_1,v_2,\cdots,v_n\}$ denotes the set of brain region nodes, and $n$ denotes the number of brain region nodes, that is, the number of ROIs. Similarly, $E=\{(v_i,v_j)\}$ denotes the set of edges between the nodes of each brain region in the graph, where $v_i,v_j \in V$ ; $E$ contains $M$ elements; thus, there are $M=C_n^2$ edges in the graph. Therefore, there exists a function relation $X: V \mapsto \mathbb{R}^m$ that assigns features to each node, and a function $L: E \mapsto \mathbb{R}^m$ that assigns features to each edge. Here, $m$ denotes the length of the node feature vector, that is, the number of time slices. Moreover, the above functions can be considered as matrices $X \in \mathbb{R}^{n \times m}$ and $L \in \mathbb{R}^{M \times m}$. Therefore, $X(v_i), i = 1, 2, \cdots, n$ as the feature vector of node $v_i$, $X_k(v_i)$ denotes the feature component of $X(v_i)$ at the $k^{th}$ time slice; $L_k\left(v_i, v_j\right) = e_{i,j}^k, i \neq j, k = 1, 2, \cdots, m$, where $e_{i,j}^k$ denotes the component of the edge feature vector $e_{i,j}$ of both nodes $v_i, v_j$ at the $k^{th}$ time slice, that is $e_{i,j} = (e_{i,j}^1, e_{i,j}^2, \cdots, e_{i,j}^m), e_{i,j} \in \mathbb{R}^m$. For the convenience of follow-up, it can be straightforwardly made to define $L\left(v_i, v_j\right) = e_{i,j}$.

For a brain region node $v_p$, its adjacent nodes are defined as all its neighboring brain regions denoted as $Ne(v_p)$, that is,

$$Ne(v_p) = \left\{ q \middle| (v_p, v_q) \in E \right\} \tag{1}$$

where $v_q$ denotes the adjacent nodes of the brain region $v_p$.

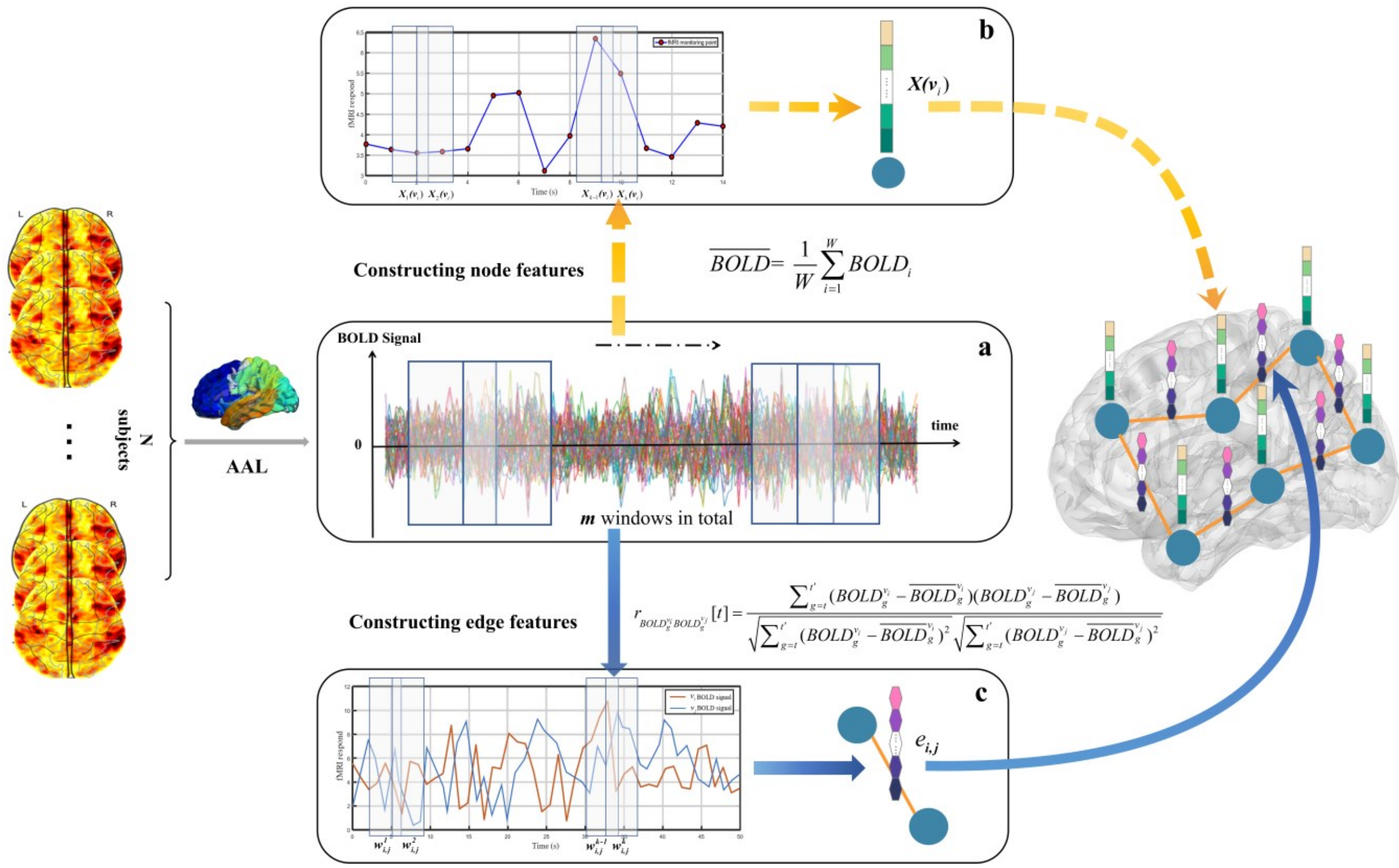

Fig. 2. Flow of DSF-BrainNet method. (a) Using the sliding window technique, the time series is sequentially divided into different time slices, (b) Using the SPA method, the node feature information under each time slice is constructed synchronously, (c) Using PCC, the constructed edge feature information under each time slice.

In Fig. 2, the rs-fMRI data of the $N$ subjects are shown on the leftmost side. The brain structure of each subject was divided into 90 ROIs using an AAL template. Thus, for each subject, the rs-fMRI data correspond to the BOLD signals of 90 ROIs. In Fig. 2(a), the entire scan time is divided into $m$ time slices using the sliding window technique, and the acquired time slices are then input into Fig. 2(b) and Fig. 2(c). Subsequently, node and edge features are constructed for the BOLD signals within each time slice using the corresponding methods. Finally, all node and edge features form a DSF-BrainNet.

### 1) Sliding window Technique

As a classical feature acquisition method, the sliding window technique has been widely applied to solve practical problems in various fields [25, 26, 27]. Therefore, in this study, sliding windows were used to construct node and edge features to exploit fully the temporal information in rs-fMRI data. Taking the rs-fMRI brain data of a subject as an example, a reasonable window width $W$ and window step $s$ were chosen for a time series of length $K$. The entire time series was divided into $m$ time slices by shifting the sliding window over the time series $K$ with a given window step $s$. The starting point of each time slice was $t \in [0, K - W]$ and the ending point was $t' = t + W$. Therefore, the total number of time slices $m$ could calculated using:

$$m = \left\lceil \frac{K - W}{s} \right\rceil + 1 \tag{2}$$

where $\lceil \ \rceil$ denotes the ceiling function.

### 2) Calculation of node features using SPA method

Because rs-fMRI is a time series consisting of multiple discrete signal points, this study extends the Piecewise Aggregation Approximation (PAA) method by applying a continuous time series to the rs-fMRI discrete time series and proposes the SPA method. After data reduction, the rs-fMRI data were organized into graph-node features.

Specifically, taking the rs-fMRI signal of any brain region in a subject, the specific mathematical procedure of SPA is expressed as follows: In each time slice, all rs-fMRI discrete signals $BOLD \in \mathbb{R}^W$ are averaged to obtain $\overline{BOLD} = \frac{1}{W}\sum_{i=1}^{W} BOLD_i$ . In total, $m$ BOLD signals were generated for each brain region in $m$ time slices, and nodal feature vectors were obtained by connecting them in time order.

### 3) Calculation of edge features using PCC

In this study, a sliding window was used to construct dFC. Specifically, PCC is used to calculate the edge feature component $e_{i,j}^k$ of the BOLD signals at each time slice for nodes $v_i$ and $v_j$ in the brain region, and the correlation coefficients at all time slices are arranged to form the edge feature vector $e_{i,j}$ .

To calculate the correlation coefficient $r\,(-1<r<1)$ for two ROIs at any time slice where the two brain region nodes are $v_i$ and $v_j$ , the PCC at the current time slice was calculated using Eq. (3):

$$r_{BOLD_g^{v_i} BOLD_g^{v_j}}[t] = \frac{\sum_{g=t}^{t'}(BOLD_g^{v_i} - \overline{BOLD}_g^{v_i})(BOLD_g^{v_j} - \overline{BOLD}_g^{v_j})}{\sqrt{\sum_{g=t}^{t'}(BOLD_g^{v_i} - \overline{BOLD}_g^{v_i})^2}\sqrt{\sum_{g=t}^{t'}(BOLD_g^{v_j} - \overline{BOLD}_g^{v_j})^2}} \tag{3}$$

where $BOLD_g^{v_i}$ and $BOLD_g^{v_j}$ denote the values of the BOLD signals of $v_i$ , $v_j$ at time point $g$ , $\overline{BOLD}_g^{v_i}$ and $\overline{BOLD}_g^{v_j}$ are the corresponding signal averages of $BOLD_g^{v_i}$ and $BOLD_g^{v_j}$ at time point $g$ , respectively. The sliding window is translated over the time series $K$ according to window step $s$ , and each translation calculates a new correlation coefficient for the current time slice, so that a total of $m$ correlation coefficients are generated between the two brain nodes at $m$ time slices. The above $m$ Pearson correlation coefficients are concatenated in time to obtain the edge feature vectors $e_{i,j}$ of $v_i$ and $v_j$ , that is, the edge feature information data between nodes $v_i$ and $v_j$ are obtained for subsequent training. The specific forms are given by Eq. (4).

$$e_{i,j} = \left\{ r_{BOLD^{v_i} BOLD^{v_j}}[1], r_{BOLD^{v_i} BOLD^{v_j}}[2], \cdots, r_{BOLD^{v_i} BOLD^{v_j}}[m] \right\}^T \in \mathbb{R}^m \tag{4}$$

## C. TemporalConv

Most graph convolution methods cannot handle multi-dimensional edge feature vectors. Those that can handle them mostly make use of fully connected operation, i.e., transforming the dimensionality of the edge feature vectors to match the node dimensionality through a linear layer (multiple linear layers form an MLP in ECConv), such as in GATConv [28] and UniMPConv [29]. Because the fully connected approach for DSF-BrainNet destroys the temporal synchronization between features, this study improved on the traditional edge convolution approach represented by ECConv by proposing TemporalConv.

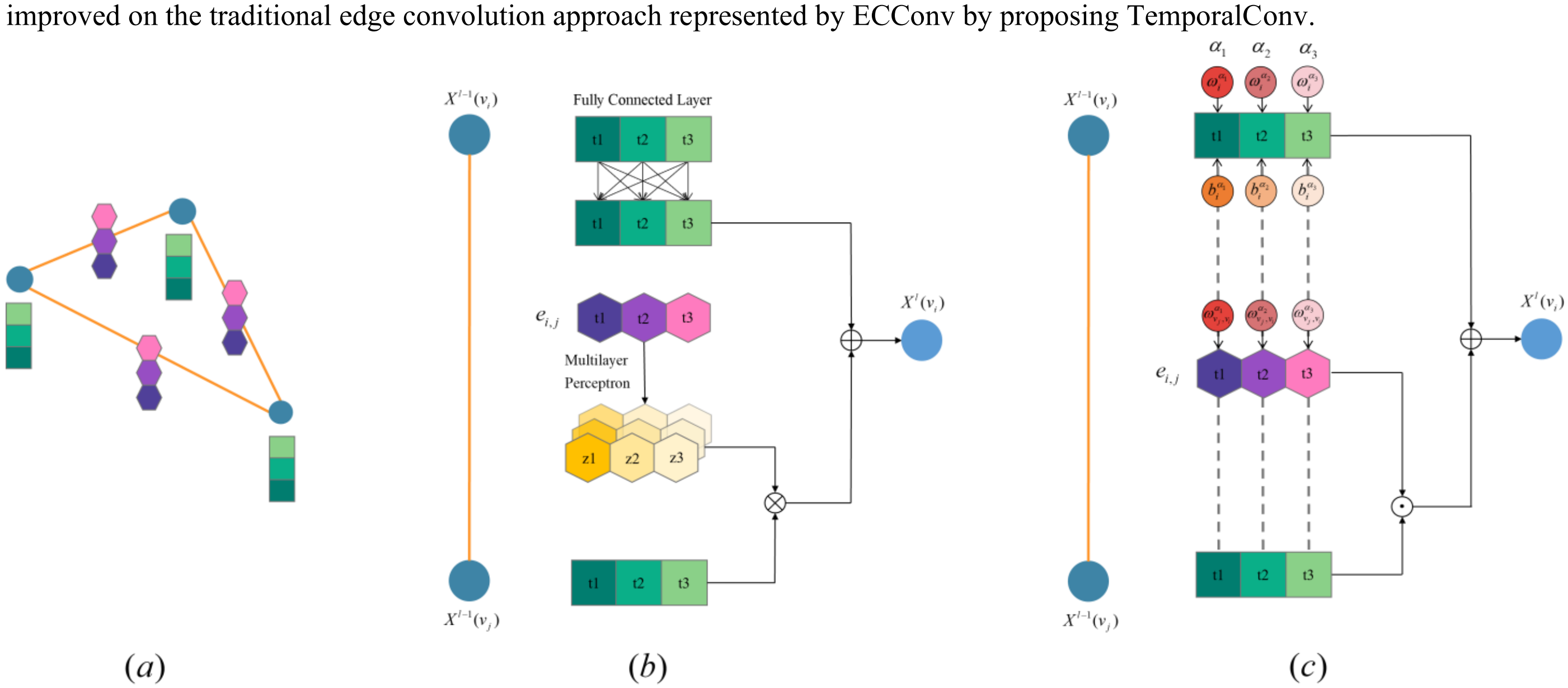

Fig. 3. Comparison of ECConv and TemporalConv convolution processes: (a) Schematic diagram of DSF-BrainNet. (b) Convolution process of applying ECConv to DSF-BrainNet. (c) Convolution process of applying TemporalConv to DSF-BrainNet.

In Fig. 3(a), the green gradient bars represent the node feature vectors and the purple gradient blocks represent the edge feature vectors, Fig. 3(b) shows the use of the ECConv method for the DSF-BrainNet constructed in this study, where $\otimes$ represents matrix multiplication; $\oplus$ represents element-wise addition, and $t1-t3$ represents different time slices of the feature vectors (that is, feature dimensions). Specifically, for the target node $X^{l-1}(v_i)$ to be convolved, ECConv inputs adjacent edge features into an MLP for dimension adjustment. Matrix multiplication is performed between the adjusted matrix and the corresponding adjacent node feature vectors. Finally, the result is added element-wise to the self-connected target node $X^{l-1}(v_i)$ to obtain the convolved target node $X^{l}(v_i)$. However, the fully connected operation in MLP and self-connection destroy the synchronous timing relationship of features; therefore, the ECConv method is not suitable for the data organization form of this study, Fig. 3(c) represents the application of TemporalConv to the DSF-BrainNet constructed in this study, where $\odot$ represents the element-wise product. Specifically, in this convolution process, each graph node in each time slice forms an independent convolution path with its neighboring nodes and edges to ensure that the features of the convolved graph nodes remain in temporal order. Also, unlike ECConv where all edge vectors share a single learnable weight in the MLP, TemporalConv sets a unique adaptive weight for each edge feature component in each time slice to facilitate the importance of all edge feature components in all time slices. This information transfer process is bidirectional as the nodes in the graph are neighbors of each other. The independent convolution process for each time slice is represented by the gray dashed line in the Fig. 3. In summary, TemporalConv preserves the temporal synchronization of features better than ECConv in the convolution process.

As the notation mentioned in Section III.B, it is assumed here that $l \in \{0,1,\cdots,l_{\max}\}$ is the layer index in a feed-forward neural network. Then there exists function relation $X^l : V \mapsto \mathbb{R}^{d_l}$ to assign features to each graph node, and function relation $L^l : E \mapsto \mathbb{R}^{d_l}$ to assign features to each edge, where $d_l$ is the feature dimension of the nodes and edge feature vectors of the $l^{th}$ layer of the neural network. Simultaneously, the above functions can also be regarded as matrices $X^l \in \mathbb{R}^{n\times d_l}$ and $L^l \in \mathbb{R}^{M\times d_l}$, where $n$ is the number of nodes in the graph, and $M$ is the number of edges in the graph. Let $\alpha \in \{\alpha_1, \alpha_2 ... , \alpha_{d_l}\}$ be the independent convolution pathway of each time slice in TemporalConv. It should be emphasized that the feature dimension of any node and edge in each time slice is one. Therefore, taking the $\alpha$ pathway in a time slice as an example, its convolved node feature value $X^l_\alpha(v_i)$ is the sum of two parts. The first part is $X^{l-1}_\alpha(v_i)$ under the influence of the adaptive weight $\omega^\alpha_i$, and learnable bias $b^\alpha_i$, and the second part is the sum of all adjacent edge feature values $L^{l-1}_\alpha(v_j, v_i)$ of the node and the corresponding adjacent node feature values $X^{l-1}_\alpha(v_j)$ under the influence of the adaptive weight $\omega^\alpha_{v_i, v_j}$, where $v_j \in Ne(v_i)$. Therefore, the TemporalConv calculation process under the $\alpha$ pathway is as follows:

$$X^l_\alpha(v_i) = \omega^\alpha_i X^{l-1}_\alpha(v_i) + b^\alpha_i + \sum\nolimits_{v_j \in Ne(v_i)} \omega^\alpha_{v_j, v_i} L^{l-1}_\alpha(v_j, v_i) X^{l-1}_\alpha(v_j) \tag{5}$$

In addition, for a graph structure constructed without edge features, Eq.(7) can be simplified as:

$$X^l_\alpha(v_i) = \omega^\alpha_i X^{l-1}_\alpha(v_i) + b^\alpha_i + \sum\nolimits_{v_j \in Ne(v_i)} \omega^\alpha_{v_j, v_i} X^{l-1}_\alpha(v_j) \tag{6}$$

### D. CategoryPool

To improve further the classification accuracy of SZ patients and analyze hemispheric lateralization abnormalities, we proposed CategoryPool. As shown in Fig. 4, this method divides the brain nodes into two categories: left and right hemispheres. After determining the total number of nodes $\gamma$ to be retained, the proportion of brain regions to be retained in these two categories ($\beta_A$, $\beta_B$) was set to investigate the effect of different retention ratios on the classification accuracy of SZ patients. Additionally, various node scoring strategies can be nested in CategoryPool pooling to filter the reserved nodes; this method mainly adopts the node scoring strategy of TopK [30] for experimental exploration.

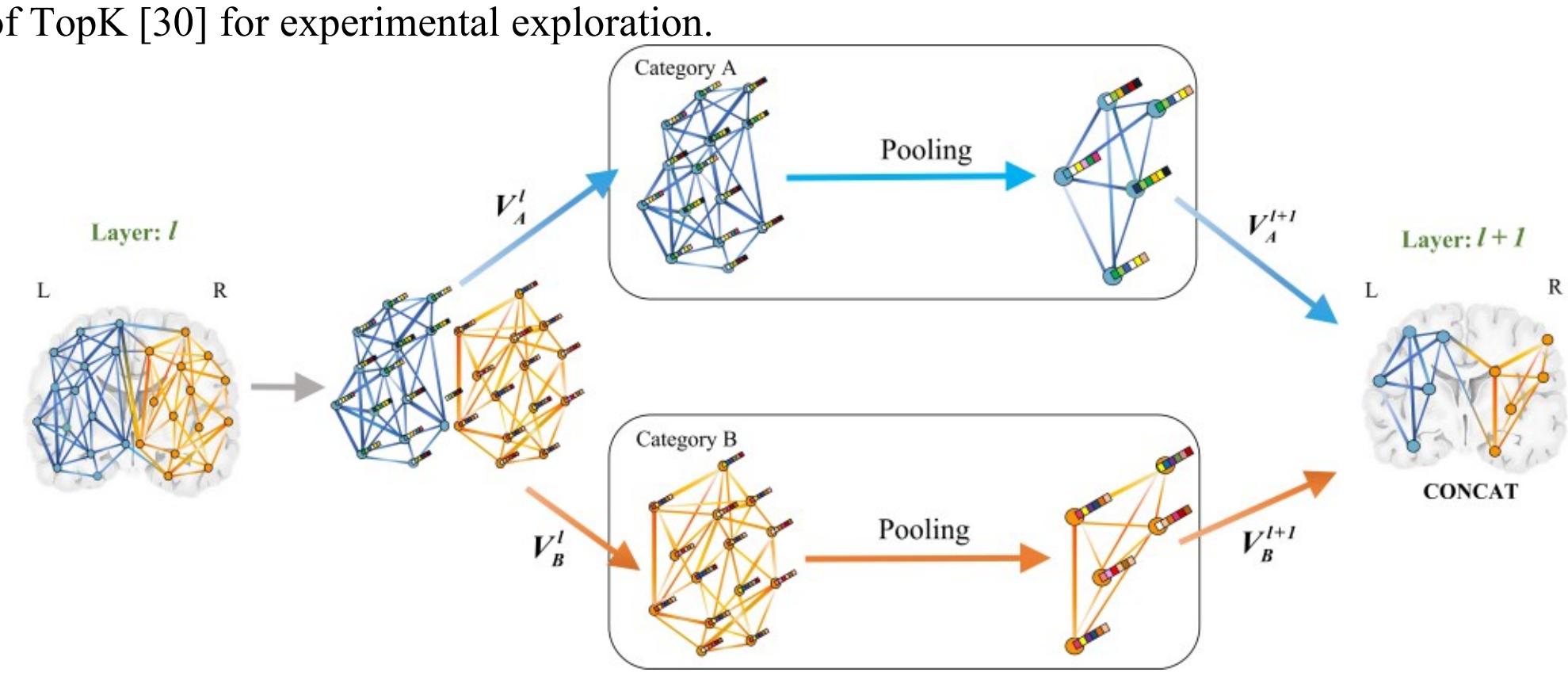

Fig. 4. CategoryPool Process Diagram.

In Fig. 4, the set of brain nodes $V^l$ of the $l^{th}$ layer neural network is firstly divided into two categories: left hemisphere $V_A^l$ and right hemisphere $V_B^l$. Secondly, intra-class pooling of nodes within the two classes using propensity parameters $\beta_A$ , $\beta_B$ respectively, is performed. In particular, the nodes in $V_A^l$ and $V_B^l$ are firstly scored and ranked adaptively and then the top $\beta_A\gamma$ and $\beta_B\gamma$ brain regions in $V_A^l$ and $V_B^l$ are extracted to form node sets $V_A^{l+1}$ and $V_B^{l+1}$ in the $l+1^{th}$ layer, respectively. Finally, the pooled $\gamma$ nodes are regrouped back into the GCN for output. Notably, $\beta_A\gamma+\beta_B\gamma=\gamma$ . Taking the brain graph of a subject as an example, $V^l=\{v_1,v_2,\cdots,v_n\}$ denotes the set of ROIs and $n$ denotes the number of ROIs. Then, all ROIs were divided into two categories: $V_A$ and $V_B$ in the left and right hemispheres, according to the AAL template, which can be denoted as

$$\begin{aligned} V_A^l &= [v_1^A, v_2^A, \cdots, v_{n_A}^A] \\ V_B^l &= [v_1^B, v_2^B, \cdots, v_{n_B}^B] \end{aligned} \tag{7}$$

where $A$ and $B$ indicates whether left and right hemisphere nodes, respectively. $n_A$ and $n_B$ denote the numbers of nodes in $V_A^l$ and $V_B^l$, respectively, and $n_A+n_B=n$. Here, a propensity parameter $\beta_i$ is defined, where $0<\beta_i<1, i=A\ or\ B$. Then, $\beta_A\gamma$ and $\beta_B\gamma$ are the numbers of brain regions to be reserved for the left and right hemisphere nodes, respectively. Finally, the left and right hemisphere nodes are pooled independently, and the output of CategoryPool is obtained by concatenating the results of the two pools.

## IV. EXPERIMENTS AND RESULTS

To evaluate the performance of the proposed Temporal-BCGCN classification model, and to discuss the hemispheric lateralization abnormalities in SZ patients, this section is divided into three parts: 1) experimental implementation and setting, 2) comparison experiment, and 3) ablation experiment. To verify the robustness of the models, the Temporal-BCGCN and comparison models were cross-validated on two cohorts 10 times with five folds, and four evaluation metrics were used to assess the performance of model: accuracy (ACC), sensitivity (SEN), specificity (SPE), and F1-Score.

### A. Experimental Implementation and Setting

Temporal-BCGCN was implemented using the PyTorch geometry and trained on an NVIDIA GeForce RTX 3080 Ti GPU. In this study, the Adam solver [31] was used; the initial learning rate was set to 0.001, and the epoch number to 500. The window width $W$ and sliding step $s$ were set to divide the time series into $m$ time slices, and then the node features and edge feature vectors were calculated for each time slice using the SPA and PCC methods, respectively. The final DSF-BrainNet was comprised of a set of graph nodes $V$ and a set of edge features $L$.

In this study, the DSF-BrainNet was fed into a GCN classifier. The classifier primarily consisted of two convolutional layers and one pooling layer. The feature dimension of the convolutional layer was $d^0=d^1=m$. The number of nodes retained in the pooling layer was $\gamma$. The proportion of preserved brain regions in the left and right hemispheres was $\beta_A$ and $\beta_B$. We designed a three layer MLP ($10\times m$, $20\times m+1$, and two neurons per layer) that took the flattened graph as the input and predicted SZ versus HC. After debugging, we set $\gamma=10$ to achieve an optimal accuracy of 83.62% on the COBRE dataset for $W=45$, $s=5$, and $\beta_A=0.9(\beta_B=0.1)$ and achieved an optimal accuracy of 89.71% on the UCLA dataset for $W=95$, $s=10$, and $\beta_A=0.7(\beta_B=0.3)$.

### B. Model Comparison

#### 1) Comparison with the baseline model:

In this study, the baseline models based on sFC and dFC were used separately for comparison. There are five main types of baseline models.

a) Traditional machine learning (ML) algorithms for sFC: Principal component analysis (PCA) + SVM.
b) Traditional DL algorithms for sFC: Convolutional neural networks (CNN)
c) Advanced static brain network deep learning analysis method for the sFC: BrainGNN [32]
d) Advanced DL algorithms for dFC: GAT, UniMP
e) New dynamic brain network deep learning analysis methods for the dFC: HFCN [33] and CRNN [12].

The kernel function of the SVM classifier was a polynomial kernel function (linear) and the PCA (n_components = 0.9) consisted of three composite convolutional layers and one fully connected classification layer. Each composite convolutional layer consisted of a convolutional layer and a maximum pooling layer. The model architectures of the BrainGNN, HFCN, and CRNN followed the setup of the original paper, and the sliding window size of the models was adjusted accordingly in this study. The GAT model consisted of two GATConv layers, one Top-kPool, and one fully connected classification layer. The output size of the fully connected classification layer was two. The UniMP model consisted of two UniMPConv layers, one Top-kPool layer, and one fully connected classification layer. The output size of the fully connected classification layer was two.

TABLE I
COMPARISON OF THE CLASSIFICATION PERFORMANCE WITH DIFFERENT BASELINE MODELS

| | COBRE(MEAN ± STD, %) | | | | UCLA(MEAN ± STD, %) | | | |
|---|---|---|---|---|---|---|---|---|
| CLASSIFIERS | ACC | SEN | SPE | F1-SCORE | ACC | SEN | SPE | F1-SCORE |
| PCA+SVM(sFC) | 78.27 ± 6.26 | 77.84 ± 12.83 | 79.19 ± 13.58 | 76.04 ± 9.69 | 80.56 ± 3.58 | 82.42 ± 13.31 | 78.28 ± 14.34 | 80.26 ± 9.64 |
| CNN(sFC) | 77.89 ± 11.43 | 79.36 ± 10.84 | 77.24 ± 11.91 | 74.54 ± 7.18 | 79.34 ± 5.71 | 76.07 ± 8.70 | 81.15 ± 14.25 | 75.56 ± 8.67 |
| BrainGNN(sFC) | 81.28 ± 8.57 | **83.45 ± 9.06** | 79.10 ± 8.65 | 79.86 ± 6.14 | 84.93 ± 6.69 | 86.53 ± 8.39 | 82.46 ± 8.86 | 83.38 ± 6.61 |
| GAT(dFC) | 80.82 ± 6.46 | 81.55 ± 10.83 | 81.86 ± 10.37 | 78.38 ± 8.49 | 80.97 ± 5.74 | 83.84 ± 8.58 | 79.98 ± 11.82 | 80.58 ± 8.47 |
| UniMPConv(dFC) | 78.11 ± 7.24 | 79.23 ± 13.68 | 81.46 ± 12.79 | 76.45 ± 10.74 | 82.45 ± 6.25 | 84.42 ± 9.39 | 82.84 ± 11.75 | 82.72 ± 6.82 |
| HFCN(dFC) | 80.83 ± 6.86 | 77.64 ± 11.26 | 83.72 ± 9.43 | 77.84 ± 8.16 | 82.84 ± 8.76 | 79.39 ± 9.92 | 85.31 ± 10.83 | 79.11 ± 9.44 |
| CRNN(dFC) | 79.29 ± 5.35 | 77.31 ± 9.55 | 81.46 ± 11.36 | 78.13 ± 8.03 | 82.67 ± 7.58 | 83.08 ± 9.85 | 80.19 ± 10.12 | 82.31 ± 8.18 |
| Temporal-BCGCN(dFC) | **83.62 ± 4.27** | 83.21 ± 7.32 | **84.07 ± 8.95** | **80.53 ± 4.71** | **89.71 ± 4.50** | **89.92 ± 8.51** | **89.56 ± 7.27** | **89.35± 5.39** |

As shown in Table I, the temporal-BCGCN model slightly outperforms the baseline methods on the COBRE dataset and significantly outperforms the baseline methods on the UCLA dataset under a one-tailed two-sample t-test of $p < 0.05$ .Specifically, the traditional ML algorithm PCA + SVM and the traditional deep learning algorithm CNN exhibit the worst results. The GAT, UniMP, HFCN, and CRNN models using dFC features perform slightly worse than BrainGNN using sFC features, whereas the temporal-BCGCN model proposed in this study outperforms BrainGNN. Therefore, although dFC contains more feature information than sFC, the performance of the model using dFC can be better than that of the advanced sFC model only if a suitable analysis framework is adopted. For example, DSF-BrainNet was proposed to build a topology for brain networks with synchronous temporal changes. TemporalConv was then proposed to maintain the synchronous time-varying features of the brain network during convolutional pooling. Maybe these reasons are why the Temporal-BCGCN model outperformed the other baseline models on both datasets. In conclusion, the proposed method is a superior tool for dynamic brain network analysis compared to other brain network analysis methods.

#### *2) Comparison with state-of-the-art methods*

This study also compared the proposed model with the current state-of-the-art research for SZ classification and diagnosis, as shown in Table II. In both datasets, Temporal-BCGCN exhibits better performance than the other state-of-the-art methods. The proposed model improves the ACC by at least 1.20% on the COBRE dataset and 6.22% on the UCLA dataset, significantly outperforming the other methods. These results demonstrate the rationality of the Temporal-BCGCN model and its superior representative learning capability.

TABLE II
COMPARISON WITH STATE-OF-THE-ART METHODS

| Methods | Datasets | Modality | HC/SZ Classification | | |
|---|---|---|---|---|---|
| | | | ACC (%) | SEN (%) | SPE (%) |
| Wang *et al.*(2019) [34] | COBRE | sMRI, rs-fMRI | 82.42 | 88.57 | 75.00 |
| Zou *et al.*(2020) [35] | COBRE, NMorphCH | rs-fMRI | 80.49 | 83.72 | 76.92 |
| Ghosal *et al.*(2021) [36] | LIBD | fMRI, Genetic data | 58.00 | 60.00 | 56.00 |
| | BARI | | 73.00 | 66.00 | 83.00 |
| Shi *et al.* (2021) [37] | COBRE | sMRI, fMRI | 83.49 | 68.69 | **93.75** |
| Wang *et al.*(2022) [6] | COBRE | sMRI, rs-fMRI | 82.42 | 88.57 | 75.00 |
| Proposed Method | COBRE | rs-fMRI | 83.62 | 83.21 | 84.07 |
| | UCLA | | **89.71** | **89.92** | 89.56 |

### *C. Ablation Experiment*

In this section, we conduct the following ablation experiments to demonstrate the importance of the three key modules SPA, TemporalConv and CategoryPool.

#### *1) Importance of SPA algorithm in DSF-BrainNet*

In this study, the node features in DSF-BrainNet were replaced by the raw rs-fMRI sequences without reduction, and the features after reduction using the PCA algorithm. It should be noted that both the raw rs-fMRI sequence and the reduction features of the SPA or PCA algorithms have temporal characteristics. As shown in Fig. 5, Temporal-BCGCN achieves the best classification ACC and F1-Score for both COBRE and UCLA when the graph node features are constructed using the SPA algorithm. The above results indicate that the dynamic synchronization feature constructed by DSF-BrainNet using the SPA algorithm is more conducive to training the Temporal-BCGCN model.

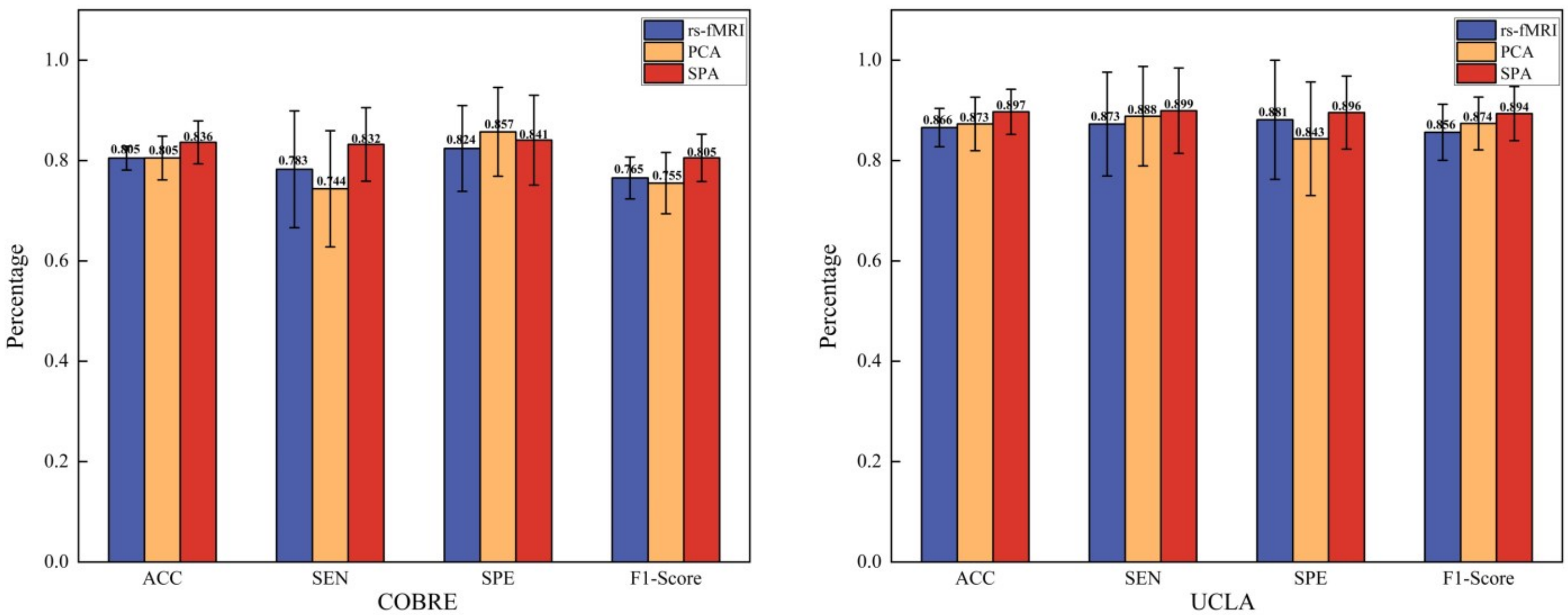

Fig. 5. SPA ablation comparison test. rs-fMRI represents the graph node features as the original rs-fMRI signal sequence.

### 2) *Superiority of TemporalConv*

To demonstrate fully the difference in the effectiveness of ECConv and TemporalConv, this study discusses the ACC of temporal-BCGCN networks based on ECConv and TemporalConv with different window widths $W$ and step sizes $s$ for the COBRE and UCLA datasets, respectively, as shown in Fig. 6. We also present the model results for ECConv, GATConv, UniMPConv, and TemporalConv under the optimal hyperparameter configuration in Table III.

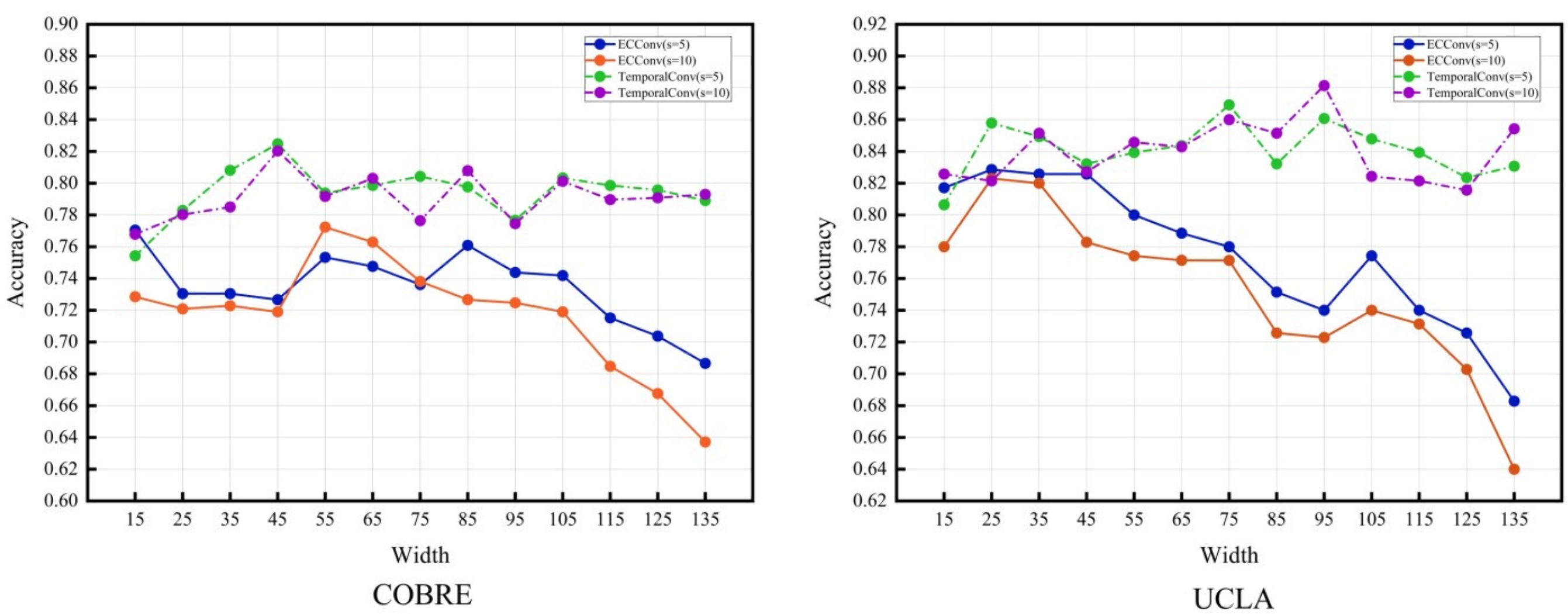

Fig. 6. ACC of Temporal-BCGCN models based on ECConv and TemporalConv with different window widths $W$ and step sizes $s$.

As shown in Fig. 6, the ACC of the Temporal-BCGCN model with TemporalConv is higher than that with ECConv, and it remains stable at a higher level than that of ECConv for different parameter settings. The ACC of the ECConv-based model decreases as the window width $W$ increases, apparently because ECConv convolves node and edge features in a fully connected manner, which results in the imperfect exclusion of redundant information. Specifically, as W increases, more time slices overlap, and more redundant information is present in the node and edge vectors. However, ECConv feeds all edge vectors into a shared MLP for fully connected operation during convolutional message passing, and the adaptive parameters of this MLP are simultaneously applied to each component of each edge vector under different time slices. As the overlap between time slices increases, the redundancy of the data corresponding to the adaptive parameters increases further, resulting in a rapid decrease in the ability of ECConv to exclude redundant information. TemporalConv introduces an exclusive adaptive parameter $\omega^{\alpha}_{v_i,v_j}$ to adjust the current edge vector components dynamically, which to a certain extent eliminates the mutual interference between time slices and reduces the redundancy of information. In summary, TemporalConv can capture the implied brain dynamics more stably when excluding overlapping information between time slices.

TABLE III
COMPARISON OF TEMPORALCONV WITH VARIOUS GRAPH CONVOLUTION STRATEGIES

| | COBRE(mean ± std, %) | | | | UCLA(mean ± std, %) | | | |
|---|---|---|---|---|---|---|---|---|
| Conv | ACC | SEN | SPE | F1-Score | ACC | SEN | SPE | F1-Score |
| ECConv | 75.24 ± 5.36 | 70.08 ± 16.37 | 80.86 ± 11.87 | 69.80 ± 8.13 | 82.86 ± 6.40 | 81.88 ± 10.90 | 81.22 ± 9.74 | 80.07 ± 7.11 |
| GATConv | 81.43 ± 7.81 | 82.19 ± 15.21 | 82.58 ± 10.49 | 78.97 ± 9.42 | 81.72 ± 4.26 | 84.76 ± 9.69 | 80.65 ± 12.92 | 81.79 ± 4.00 |
| UniMPConv | 78.67 ± 5.86 | 75.82 ± 14.39 | **86.45 ± 11.16** | 73.36 ± 11.49 | 83.71 ± 5.27 | 85.55 ± 10.88 | 83.61 ± 10.30 | 83.29 ± 5.68 |
| TemporalConv | **83.62 ± 4.27** | **83.21 ± 7.32** | 84.07 ± 8.95 | **80.53 ± 4.71** | **89.71 ± 4.50** | **89.92 ± 8.51** | **89.56 ± 7.27** | **89.35± 5.39** |

As shown in Table III, TemporalConv exhibits the best overall performance on the COBRE and UCLA datasets under the optimal hyperparameter configuration. In particular, compared with the other models, all metrics are optimal and exceed 80%, except for the SEN of COBRE, which is lower than that of the UniMPConv model.

### 3) Superiority of CategoryPool

In this study, CategoryPool was used under the Top-kPool, SortPool [38] and SAGPool [39] pooling strategies. As shown in Table IV, the average accuracies of the models on the COBRE and UCLA datasets improved by 2.32% and 4.14% respectively when each pooling method was nested using CategoryPool than when used alone. Simultaneously, the standard deviations of the models nested in CategoryPool converged; that is, the model quality was more stable. In summary, CategoryPool can improve the performance of various pooling strategies for specific problems. The reason for this improvement may be that, for rs-fMRI data, pooling the left and right hemispheres into different categories not only adapts to the characteristics of the data, but also captures some key features.

TABLE IV
COMPARISON OF THE USE OF VARIOUS POOLING STRATEGIES BASED ON CATEGORYPOOL

| | COBRE(mean ± std, %) | | | | UCLA(mean ± std, %) | | | |
|---|---|---|---|---|---|---|---|---|
| Pooling | ACC | SEN | SPE | F1-Score | ACC | SEN | SPE | F1-Score |
| Top-kPool | 82.48 ± 3.79 | 83.17 ± 8.64 | 83.71 ± 8.20 | 79.45 ± 5.71 | 88.14 ± 5.20 | 88.75 ± 8.07 | 89.45 ± 8.34 | 87.36 ± 5.90 |
| CategoryPool(Top-k) | **83.62 ± 4.27** | **83.21 ± 7.32** | **84.07 ± 8.95** | **80.53 ± 4.71** | **89.71 ± 4.50** | **89.92 ± 8.51** | **89.56 ± 7.27** | **89.36± 5.39** |
| SortPool | 79.62 ± 4.07 | 69.59 ± 9.47 | **87.55 ± 6.61** | 73.91 ± 6.11 | 81.72 ± 6.03 | 88.75 ± 12.09 | 77.42 ± 12.79 | 81.97 ± 6.56 |
| CategoryPool(Sort) | **80.19 ± 3.52** | **72.89 ± 8.75** | 86.78 ± 7.51 | **75.55 ± 4.79** | **89.71 ± 4.41** | **92.85 ± 9.64** | **87.82 ± 9.07** | **89.11 ± 6.04** |
| SAGPool | 79.05 ± 4.41 | 66.02 ± 9.60 | 90.02 ± 6.95 | 72.50 ± 4.79 | 83.43 ± 5.54 | 85.36 ± 11.84 | 84.41 ± 12.93 | 82.76 ± 6.86 |
| CategoryPool(SAG) | **84.29 ± 4.15** | **74.75 ± 7.23** | **92.37 ± 9.19** | **79.98 ± 5.68** | **86.28 ± 3.89** | **88.50 ± 9.03** | **86.90 ± 11.21** | **85.74 ± 5.32** |

## V. DISCUSSION

To explore the hemispheric laterality abnormality in SZ patients, the brain was divided into two categories: left and right hemispheres, then pooled with CategoryPool, and $\beta_A$ and $\beta_B$ were used to explore the lateralization abnormality of the brain and its disease biomarkers. In this section, the lateralization abnormalities in the brain of sperm patients will be explored specifically by discussing the existence of hemispheric lateralization abnormalities in SZ patients and providing an analysis of these abnormalities in SZ patients.

### A. Existence of hemispheric lateralization abnormalities in SZ patients

TABLE V
ORDER OF RELATIVE IMPORTANCE OF PATHOLOGICAL TOP10 BRAIN REGIONS WHEN $\beta_A$ IS EQUAL TO 0.5

| COBRE | | UCLA | |
|---|---|---|---|
| ROI abbr | Brain Network | ROI abbr | Brain Network |
| **SFGmed.L** | DMN | **SFGmed.L** | DMN |
| **PCUN.R** | DMN | ITG.L | LN |
| LING.R | VN | **PCUN.R** | DMN |
| **SOG.L** | VN | CAL.L | VN |
| HIP.L | - | **FFG.L** | VN |
| STG.L | SMN | MTG.R | DMN |
| PreCG.L | DAN | SFGdor.R | DMN |
| LING.L | VN | ACG.L | FPN |
| **FFG.L** | VN | DCG.L | VAN |
| PCG.R | FPN | **SOG.L** | VN |

DMN: default-model network, DAN: dorsal attention network, FPN: frontoparietal network, LN: limbic network, SMN: somatosensory network, VAN: ventral attention network, VN: visual network, -: not in the cortical network.

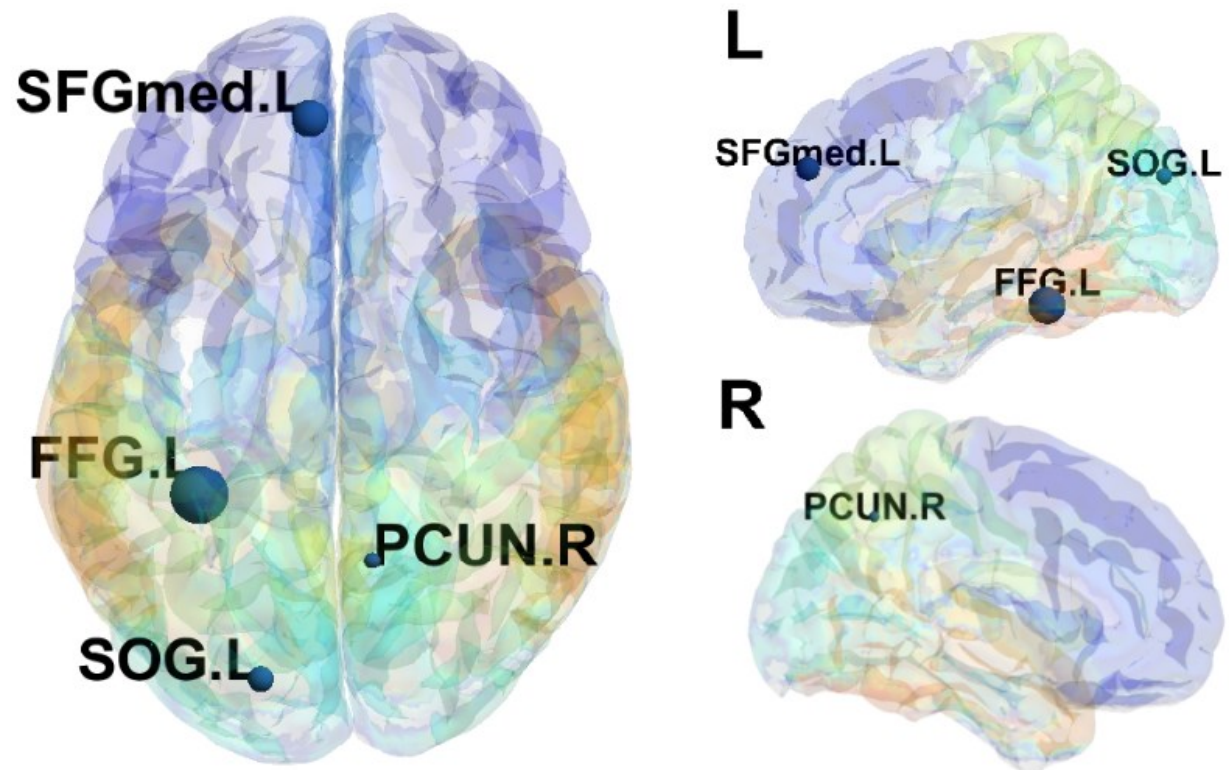

Fig. 7. Potential disease biomarkers under $\beta_A = 0.5(\beta_B = 0.5)$ (as shown in Table V, the brain regions that appear in the importance sequences of both data sets are regarded as potential disease biomarkers).

To test whether there is abnormal lateralization in the brains of SZ patients, we firstly set the same status between the left and right hemispheres, namely, $\beta_A = 0.5(\beta_B = 0.5)$. Secondly, during pool training, adaptive scores were introduced to describe the importance of preserving the brain areas. Finally, the adaptive scores of the brain regions retained 10 times 5-fold cross-validations were accumulated and ranked; the top 10 brain regions are shown in Table V, where the number of left hemisphere regions is seven in both datasets, significantly exceeding the right hemisphere. This finding also suggests that pathological changes in the left hemisphere of SZ patients may be more favorable for model classification training. In summary, there were more group-differentiated lesions in the left hemisphere of SZ patients. Moreover, brain regions that appeared simultaneously in both datasets were regarded as potential disease biomarkers (bold in Table V), as shown in Fig. 7. Specifically, under the condition of equal status between the left and right hemispheres, the potential disease biomarkers identified in this study were located in the DMN (PCUN.RandSFGmed.L) and the VN (SOG. LandFFG.L).

## *B. Analysis of hemispheric lateralization abnormalities in SZ patients*

TABLE VI

ORDER OF RELATIVE IMPORTANCE OF TOP10 BRAIN REGIONS UNDER OPTIMAL ACCURACY

| COBRE | | UCLA | |
|---|---|---|---|
| ROI abbr | Brain Network | ROI abbr | Brain Network |
| SOG.L | VN | LING.L | VN |
| FFG.L | VN | **SFGmed.L** | **DMN** |
| **SFGmed.L** | **DMN** | **IOG.L** | **VN** |
| TPOmid.L | LN | ACG.L | VAN |
| MOG.L | VN | **PreCG.L** | **DAN** |
| **CAL.L** | **VN** | **CAL.L** | **VN** |
| CUN.L | VN | MTG.L | LN |
| **IOG.L** | **VN** | ITG.L | LN |
| STG.L | SMN | DCG.L | VAN |
| **PreCG.L** | **DAN** | PoCG.L | SMN |

DMN: default-model network, DAN: dorsal attention network, LN: limbic network, SMN: somatosensory network, VAN: ventral attention network, VN: visual network, -: not in the cortical network.

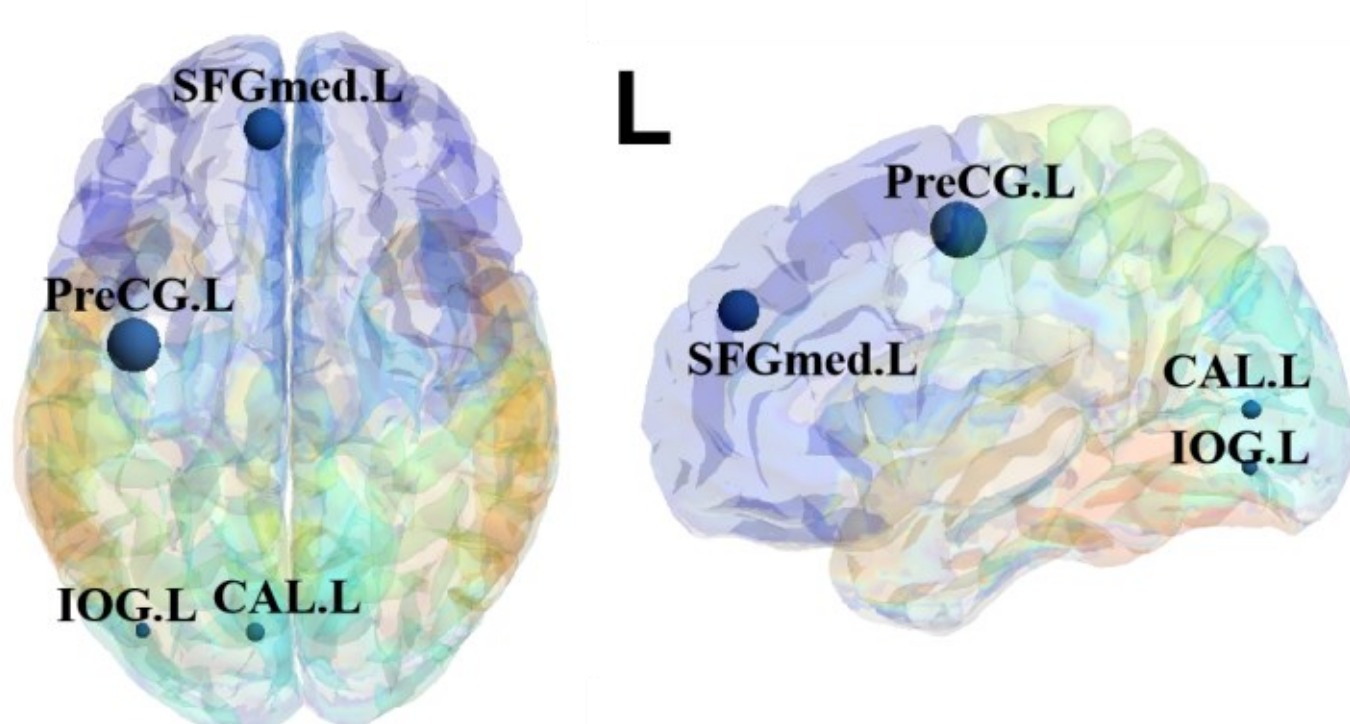

Fig. 8. Potential disease biomarkers with optimal accuracy (as shown in Table VI, the brain regions that appear in the importance sequences of both data sets are regarded as potential disease biomarkers).

To analyze further the abnormal lateralization of the left hemisphere in SZ patients, under the COBRE and UCLA datasets, the model achieved the highest accuracy when $\beta_A = 0.9(\beta_B = 0.1)$ and $\beta_A = 0.7(\beta_B = 0.3)$ (i.e., when it was more inclined to choose left brain features), respectively. The importance of the lesions in the brain regions corresponding to the above values is listed in Table VI (top 10). Under both datasets, the top 10 brain regions were located in the left hemisphere, which further verified the conclusion drawn in Part A of this section. Additionally, brain regions that appeared simultaneously in both datasets were regarded as potential disease biomarkers (bold in Table VI), as shown in Fig. 8. Specifically, the disease biomarkers identified in this study were located in the VN (IOG. LandCAL. L), DAN (PreCG. L), and DMN (SFGmed. L) with the highest accuracy.

In summary, the results of the present study suggest that the left hemisphere of the brain in SZ patients seems to play a more critical role in classification prediction, which may suggest more severe lesions in the left hemisphere of the brain in SZ patients. Both Tables V and VI show that accurate classification of SZ can be achieved based on the functional characteristics between the lower order perceptual networks (VN and SMN) and the higher order networks (DMN and LN). Indeed, it has been demonstrated that SZ involves many brain regions and is considered a widespread disconnected brain disease [40, 41, 42]. Meanwhile, the presence of functional abnormalities in DMN brain regions was shown to be associated with reduced interhemispheric FC in the left intrahemisphere [4]. Xie *et al.* [18] found that the syrinx gyrus in DMN of schizophrenic patients exhibited greater rs-fMRI signal variability and that this variability tended to be lateralized to the left.

More importantly, the SFGmed.L, located in the DMN network, was labeled as an important potential disease biomarker in Tables V and VI. Moreover, studies have identified the importance of this brain region in SZ. For example, Zhang *et al.* [19] observed significant hemispheric effects in 10 brain regions, including the precentral gyrus and medial part of the superior frontal gyrus, as well as the precentral gyrus and medial part of the superior frontal gyrus, which showed a left-leaning predominance in terms of regional efficiency. Pan *et al.* [43] found a significant reduction in connectivity at the whole-brain level in SZ patients whose working memory capacity was positively correlated with the strength of connections in SFGmed.L-SFGmed.R and SFGmed.L-ACG.R. Furthermore, consistent with the sensory gating hypothesis, the present study found significant predictive precision in the lower order perceptual system (occipital and primary sensorimotor cortex), reflecting dysregulation of the primary perceptual system in these two datasets [44]. Higher order brain regions, such as the prefrontal cortex, also showed good discrimination, which may be related to the executive dysfunction of the patients.

## VI. Conclusion

In summary, in the future, we will develop more targeted deep learning functional modules based on the characteristics of medical image data itself to realize fully the potential of deep learning methods in medical image processing. In this study, a powerful and advanced model for dynamic brain network analysis, Temporal-BCGCN, was employed for computer-aided diagnosis and hemispheric lateralization analysis of SZ patients. The proposed model includes the following: (1) a unique dynamic functional brain network module DSF-BrainNet; (2) a revolutionary convolution method TemporalConv; and (3) a novel pooling strategy CategoryPool. The proposed model yielded average accuracies of 83.62% and 89.71% for the COBRE and UCLA datasets, respectively, exceeding those of the baseline models and other state-of-the-art methods. This study showed that the lower order perceptual system and higher order network regions in the left hemisphere are more severely dysfunctional than those in the right hemisphere in SZ. It also confirmed the significance of the left medial superior frontal gyrus (SFGmed.L) in SZ. In conclusion, Temporal-BCGCN provides a promising means of restoring the spatiotemporal characteristics of cerebral neurodynamics and identifies the lateralization abnormalities of SZ patients. In the future, we will extend the use of TemporalConv to explore further the patterns of abnormal activation under each time slice of the brain.